\pdfoutput=1
\documentclass[aps,twocolumn, floatfix, prb]{revtex4-1}
\usepackage{graphicx}
\DeclareGraphicsExtensions{.pdf}
\usepackage{amsmath,amssymb,bbold,bm,color}
\usepackage{float}
\usepackage{epstopdf}
\usepackage{siunitx}

\newcommand{\bk}{{\bm k}}

\newcommand{\br}{{\bm r}}

\newcommand{\bE}{{\bm E}}

\newcommand{\ba}{{\bm a}}
\newcommand{\bx}{{\bm x}}
\newcommand{\by}{{\bm y}}
\newcommand{\bI}{{\bm I}}

\newcommand{\bj}{{\bm j}}

\newcommand{\cC}{{\cal C}}

\newcommand{\cT}{{\cal T}}

\newcommand{\cI}{{\cal I}}
\newcommand{\cV}{{\cal V}}
\newcommand{\cR}{{\cal R}}

\newcommand{\bee}{\begin{equation}}
\newcommand{\ee}{\end{equation}}

\newcommand{\ket}[1]{| #1 \rangle}

\def\sgn{\mathop{\rm sgn}\nolimits}

\begin{document}

\title{Chern insulators for electromagnetic waves in electrical circuit networks}

\author{Rafael Haenel}
\affiliation{Department of Physics and Astronomy, University of
British Columbia, Vancouver, BC, Canada V6T 1Z1}
\affiliation{Quantum Matter Institute, University of British Columbia, Vancouver BC, Canada V6T 1Z4}
\author{Timothy Branch}
\affiliation{Department of Physics and Astronomy, University of
British Columbia, Vancouver, BC, Canada V6T 1Z1}
\affiliation{Quantum Matter Institute, University of British Columbia, Vancouver BC, Canada V6T 1Z4}
\author{Marcel Franz}
\affiliation{Department of Physics and Astronomy, University of
British Columbia, Vancouver, BC, Canada V6T 1Z1}
\affiliation{Quantum Matter Institute, University of British Columbia, Vancouver BC, Canada V6T 1Z4}

\begin{abstract} 
Periodic networks composed of capacitors and inductors have been demonstrated to possess
topological properties with respect to incident electromagnetic waves. Here, we
develop an analogy between the mathematical description of waves
propagating in such networks 
and models of Majorana fermions hopping on a lattice. Using this
analogy we propose simple electrical network architectures that
realize  Chern insulating phases for electromagnetic waves. Such Chern insulating networks have a bulk gap for a range of signal frequencies that is easily tunable and exhibit topologically protected chiral
edge modes that traverse the gap and are robust to perturbations. The requisite time reversal symmetry breaking is achieved
by including a class of weakly dissipative Hall resistor elements
whose physical  implementation we describe in detail.
\end{abstract}

\date{\today}
\maketitle

\section{Introduction}
Topological states of matter in electronic systems exhibit topologically non-trivial bulk band structures accompanied by protected edge or
surface modes \cite{Hasan.2010,Qi.2011,Franz2013}. More generally, the insights gained from the
study of electrons in crystalline solids with non-trivial topology can be applied to
any physical system whose degrees of freedom are governed by a wave equation.
If bulk solutions of the
wave equation do not exist in some range of frequencies, the
system may be viewed as insulating for these frequencies and may
in addition possess topologically protected propagating modes at its boundary. 
This realization has led to a theoretical study and physical
implementation of a wide variety of periodic systems
in which topological properties analogous to electronic topological
insulators, superconductors and semimetals are manifest. 
Most prominent examples of
these efforts include photonic \cite{Haldane.2008,Lu.2014}, acoustic \cite{Prodan2009, Yang.2015},
mechanical \cite{Kane2014, Huber.2016,Barlas2018}, polaritonic \cite{Karzig.2015}, and electrical
systems \cite{Ningyuan.2015, Lee.2018, Zhao.2018, Luo2018}.

In the present work, we focus on the latter class of topological systems, more specifically,
periodic networks comprised of inductors, capacitors and resistors. These structures,
also referred to as topoelectrical circuits \cite{Lee.2018}, have been demonstrated to
possess topological properties with respect to the incident electromagnetic (EM)
wave signals. In close analogy to electronic tight-binding models, 
various circuit models realizing classical analogs of
quantum spin Hall states \cite{Ningyuan.2015, Zhu.2018}, Dirac and Weyl semimetals\cite{Lee.2018,lu2018probing} and higher order topological insulators\cite{Ezawa2018, Ezawa2018a,Imhof2018} have been 
proposed and some of them have been experimentally characterized.

Conspicuously absent from this list is the Chern insulator -- the analog of the most 
basic electronic topological phase, the quantum Hall insulator in two
dimensions \cite{Haldane.1988}. The reason is simple: networks composed of capacitors and inductors
are governed by Maxwell equations which are fundamentally invariant
under the time reversal operation $\cT$. A Chern insulator, on the other hand, requires broken
$\cT$ symmetry. 

We note that ordinary resistors in a LC network cause dissipation and
therefore break $\cT$.  This to some extent hinders the comparison to
isolated quantum systems where dynamics are unitary. More importantly, 
$\cT$-breaking produced by purely dissipative dynamics does not help in creating a Chern insulator. 
On the other hand dissipative networks can provide useful examples of
systems studied in the rapidly advancing field of 
non-Hermitian quantum mechanics \cite{Shen.2018, Gong2018, Luo2018nonhermitian, Moiseyev2011}.

Here, we circumvent this problem by employing a class of
weakly dissipative Hall resistors. These are linear circuit elements whose
voltage response to a longitudinal current is predominantly
transverse. An ideal Hall resistor
introduces strong $\cT$ breaking into the circuit without significant dissipation
and thus enables construction of the Chern insulator.

The class of EM Chern insulators we introduce here has a bulk gap for 
EM waves in a range of frequencies but exhibits chiral propagating edge modes 
that are gapless and traverse
the gap. The edge modes are topologically protected by a non-zero Chern number
defined by the bulk band structure of the network and are robust against any
imperfections in the network that do not close the gap. 

We discuss
several specific network architectures that lead to the physics described
above. We do this by taking advantage of a novel mapping that connects
the dynamics of a certain class of periodic RLC networks to Hermitian Bloch Hamiltonians 
describing Majorana fermions in a
crystal lattice. Such Hamiltonians are well known to possess
Chern-insulating phases. While the possibility of non-trivial topological structure of Kirchhoff's equations
has been previously recognized, it is usually discussed in terms of admittance bands or mapped onto 
non-Hermitian eigenvalue problems \cite{Ningyuan.2015, Lee.2018}. The description developed in this
work offers a more direct analogy to crystalline solids and thus a
more transparent physical interpretation in terms of well understood
topological band theory\cite{Hasan.2010,Qi.2011,Franz2013}.

The key physical element required in the realization of our EM
Chern insulator architecture is the Hall resistor. A natural
implementation of the Hall resistor relies 
on the classical Hall effect in a clean metal or doped 
semiconductor film in applied perpendicular magnetic
field $B_\bot$ with galvanic contacts. We will see that this simplest
realization does not quite work and discuss an alternate setup with
capacitive couplings.
We also discuss implementations where near-ideal Hall
response can be simulated using simple circuits with active elements: operational
amplifiers. We conclude that these offer the most practical route
towards the realization of Chern insulating networks that would function at room temperature and use only
ready-made components.

\begin{figure*}[htb]
\includegraphics[width=17cm]{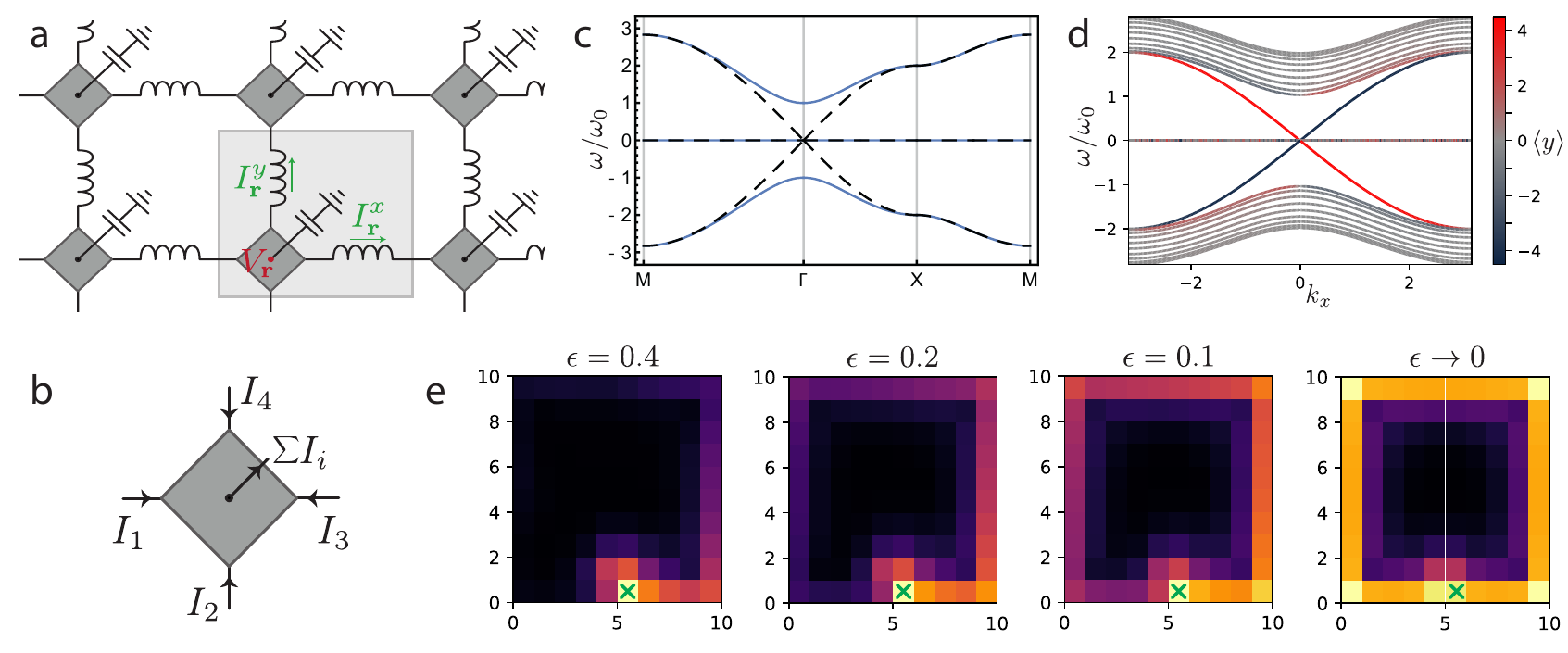}
\caption{
  (a) Square RLC lattice toy model realizing the Chern insulator
  for EM waves. A unit cell is marked by by grey background.
  (b) The Hall resistor element with four
  side terminals and one central terminal. 
  (c)  Bulk band structure of
  the network depicted in panel a. The dashed line corresponds to $\gamma=R_H\sqrt{C/L}=0$ while the solid line corresponds to
  $\gamma=0.25$ which gives a gap $\Delta=\omega_0$, where $\omega_0=1/\sqrt{LC}$. 
  (d) Spectrum on
  a strip of width $W=10$ and open boundary conditions along $y$ 
  for $\gamma=0.25$. Boundary 
  conditions are chosen as indicated in Fig.\ \ref{fig:boundaries}(b). The colorscale
  indicates the average distance $\langle y\rangle$ measured from the center of
  the strip of the eigenstate belonging to the eigenvalue
  $\omega_k$. The states inside the bulk gap $\Delta$ are
  localized near the opposite edges of the system. 
  (e) Voltage response
  $V^{\rm resp}_\br(\omega)$ 
induced by a current with in-gap frequency $\omega=\Delta/2$ injected at
a node marked by green cross of the $10\times 10$ network with $\gamma=0.25$ for
various values of the dissipative resistance $R$ characterized by
parameter $\epsilon=R/R_H$. }
\label{fig1}
\end{figure*}
\section{Chern insulators from RLC networks}


\subsection{General setup and a toy model}

The simplest RLC network capable of  exhibiting non-trivial topology
is depicted in Fig.\ \ref{fig1}(a). It consists of an array of 5-terminal Hall elements, 
denoted by grey diamonds, arranged in a square lattice. The central
terminal of each Hall element  
is connected to ground via a capacitor $C$ while the side terminals connect to 
neighboring Hall resistors through inductors $L$.

The 5-terminal Hall element is characterized by its resistance 
tensor $\hat{R}$, defined by the relation
\begin{equation}\label{rho}
\begin{pmatrix} V_1 \\ V_2 \\ V_3 \\ V_4 \end{pmatrix} =
\begin{pmatrix} 
R_1 & R_4 & R_3 & R_2 \\
R_2 & R_1 & R_4 & R_3 \\
R_3 & R_2 & R_1 & R_4 \\
R_4 & R_3 & R_2 & R_1 
\end{pmatrix}
\begin{pmatrix} I_1 \\ I_2 \\ I_3 \\ I_4 \end{pmatrix}.
\end{equation}
Here, the voltages $V_i$ are measured with respect to the
central terminal and the directionality of currents $I_i$ is indicated
in  Fig.\ \ref{fig1}(b). We note that Eq. (\ref{rho}) is the most general 
parametrization of $\hat{R}$ under 4-fold rotational symmetry.

The description of the EM signal propagating through the circuit requires
the definition of three dynamical variables: voltage $V_\br(t)$ across each capacitor, 
and two currents
$I^x_\br(t)$ and  $I^y_\br(t)$ flowing through the inductors in
each unit cell labeled by vector $\br$. They are denoted by red and green labels in
Fig.\ \ref{fig1}(a), respectively. Then, Kirchhoff laws yield the following coupled 
system of linear differential equations:
\begin{equation}\label{kir1}
\begin{split}
	C{\frac{\partial V_\br}{\partial t}} &= I^x_{\br-\hat{\bx}}+ I^y_{\br-\hat{\by}}-I^x_{\br}- I^y_{\br}, \\
V_\br-V_{\br+\hat{\bx}} &= L{\frac{\partial
I^x_\br}{\partial t}}+[\hat{R}\cdot\bI_\br]_3-[\hat{R}\cdot\bI_{\br+\hat{\bx}}]_1,
\\
V_\br-V_{\br+\hat{\by}} &= L{\frac{\partial
I^y_\br}{\partial t}}+[\hat{R}\cdot\bI_\br]_4-[\hat{R}\cdot\bI_{\br+\hat{\by}}]_2.
\end{split}
\end{equation}
The first of these equations expresses current conservation for
each Hall element and the remaining two relate the voltage
differences between neighboring unit cells to the corresponding
currents through the usual constitutive relations for inductors and
resistors. $\bI_\br=(I^x_ {\br-\hat{\bx}},I^y_
{\br-\hat{\by}},-I^x_\br,-I^y_\br)^T$
is a vector of currents flowing into the Hall resistor at
position $\br$.

\begin{figure}[htb]
  \centering
  \includegraphics{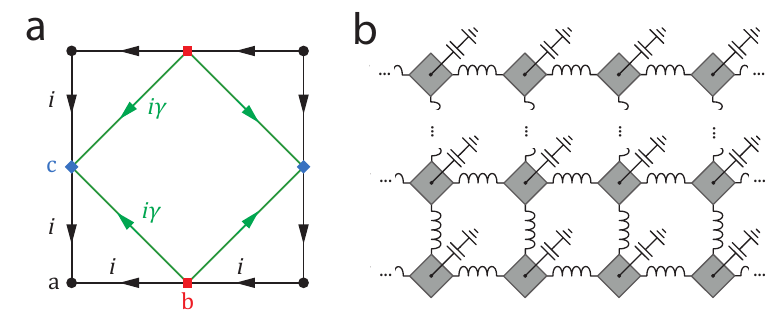}
  \caption{(a) Effective Majorana tight-binding model corresponding to the RLC network toy model with
  ideal Hall elements. Tunneling matrix elements between sublattices $a,b,c$ are labeled by straight lines, 
  arrows indicate directionality. Here, $\gamma=R_H\sqrt{C/L}$. (b) Sketch of boundary conditions used for calculations in the strip geometry.}
  \label{fig:boundaries}
\end{figure}
We begin by considering the case of a non-resistive network, i.e.\ $\hat{R}=0$. 
Then  Eqs.\
(\ref{kir1})  exhibit invariance under $\cT$ which sends $t\to -t$ and
reverses all currents, $(I^x_\br,I^y_\br)\to (-I^x_\br,-I^y_\br)$. In addition, because
voltages and currents are by definition real-valued,  Eqs.\
(\ref{kir1}) are trivially invariant under complex conjugation. 

Eqs.\ (\ref{kir1}) can be recast in the form of a
Schr\"odinger equation $i\partial_t\ket{\Psi}=\hat{H}\ket{\Psi}$ with the wavefunction
$\ket{\Psi}$ containing voltages and currents and $\hat{H}$ the Hermitian Hamiltonian
matrix. We can further exploit translational invariance of the network
by expanding currents and voltages in terms of plane waves
\begin{equation}\label{kir2}
\begin{split}
V_\br(t)&=\sum_{\bk} e^{i(\omega t-\bk\cdot\br)}\cV_\bk/\sqrt{C},\\
I^\alpha_\br(t)&=\sum_{\bk} e^{i(\omega t-\bk\cdot\br)}\cI^\alpha_\bk/\sqrt{L},
\end{split}
\end{equation}
where $\alpha=x,y$ and the rescaling is made for convenience. 
Equations\ (\ref{kir1}) reduce to a $3\times 3$ Hermitian eigenvalue problem
$H_{\bk} \ket{\psi} = \omega_{\bk} \ket{\psi}$, where
\begin{equation}\label{kir3}
\ket{\psi}=
\begin{pmatrix} \cV_\bk \\ \cI^x_\bk \\  \cI^y_\bk \end{pmatrix} 
\,,
\quad
H_{\bk}=
{\frac{1}{\sqrt{LC}}}
\begin{pmatrix} 
0 & \Gamma_x & \Gamma_y  \\
\Gamma_x^*  & 0 & 0\\
\Gamma_y^*  & 0 & 0
\end{pmatrix}
,
\end{equation}
and $\Gamma_\alpha=i(1-e^{ik_\alpha})$. 

$H_{\bk}$ is formally
identical to a tight-binding model of Majorana fermions on the
Lieb lattice. The effective electronic unit cell with imaginary hopping 
parameters is sketched in Fig.\ \ref{fig:boundaries}(a). 
The correspondence with Majorana as opposed to
complex fermions follows from the fact that the original wave equation
(\ref{kir1}) is purely real-valued as is the time-domain Schr\"odinger
equation for Majorana fermions\cite{Alicea2012,Beenakker2013,Elliott2015}.  We discuss this correspondence more fully in Appendix A.

The spectrum of $H_\bk$  consists of one zero mode $\omega_\bk=0$, and two
non-zero eigenvalues of the form
\begin{equation}\label{kir4}
\begin{split}
	\omega_\bk&=\pm{\frac{1}{\sqrt{LC}}}\sqrt{|\Gamma_x|^2+|\Gamma_y|^2} \\
	&=\pm{\frac{2}{\sqrt{LC}}}\sqrt{\sin^2{(k_x/2)}+\sin^2{(k_y/2)}}.
\end{split}
\end{equation}
It can be checked that the states belonging to the $\omega_\bk=0$
eigenvalue correspond to static patterns of currents in the network
consistent with current conservation and zero voltages. These will be
damped in the presence of arbitrary resistance and are of no interest
to us. The two branches in Eq.\ (\ref{kir4}) define the propagating
modes of the system. They are gapless and linearly dispersing near
$\bk=0$, as illustrated in Fig.\ \ref{fig1}(c). 
  Only the positive-frequency branch is physical; the negative
branch appears because the ansatz in Eq.\  (\ref{kir2}) permits complex-valued
solutions while voltages and currents are strictly real.

In the Bloch Hamiltonian formulation 
time reversal symmetry $\cT$ and charge conjugation symmetry $\cC$ 
may be expressed as
\begin{equation}\label{kir5}
\begin{split}
\cT: & \ \ \ \Theta H_\bk^* \Theta^{-1}= H_{-\bk}, \\
\cC:& \ \ \ H_\bk^* = -H_{-\bk}.
\end{split}
\end{equation}
with $\Theta={\rm diag}(1,-1,-1)$. 
Both $\cT$ and $\cC$ square to $+1$ and thus define the BDI class in the
Altland-Zirnbauer classification \cite{Altland.1997}. In two spatial dimensions class BDI
supports only topologically trivial gapped phases \cite{Schnyder.2008}. Therefore, we must
break time reversal symmetry to enable a topological phase in this system. 
(The $\cC$ symmetry
derives from real-valuedness of Eq.\ (\ref{kir1}) and therefore, like
the analogous symmetry present  in
a generic superconductor, cannot be broken by a physical perturbation.) 
When $\cT$ is broken, the system belongs to class D which has an
integer topological classification in $d=2$. The corresponding
topological invariant is the Chern number $c$ and its non-zero values
label distinct Chern insulating phases. 

To proceed, we now include a non-zero resistance tensor
defined by Eq.\ (\ref{rho}). The Bloch Hamiltonian describing the
network becomes  
\begin{equation}\label{h1}
H_\bk =
{\frac{1}{\sqrt{LC}}}
\begin{pmatrix} 
0 & \Gamma_x & \Gamma_y  \\
\Gamma_x^*  & L^x_\bk & M_\bk+N_\bk\\
\Gamma_y^*  & M_\bk^*-N_\bk^* & L^y_\bk
\end{pmatrix},
\end{equation}
with 
\begin{equation}\label{kir6}
\begin{split}
	L^\alpha_\bk&=2i\sqrt{\frac{C}{L}}(R_3\cos{k_\alpha}-R_1),\\ 
	M_\bk&= -{\frac{i}{2}}\sqrt{\frac{C}{L}}(R_4-R_2)(1+e^{ik_y})(1+e^{-ik_x}), \\
	N_\bk&= -{\frac{i}{2}}\sqrt{\frac{C}{L}}(R_4+R_2)(1-e^{ik_y})(1-e^{-ik_x}).
\end{split}
\end{equation}
Time-reversal is explicitly broken whenever $\hat{R}$ is non-zero. 
We observe that the Hamiltonian (\ref{h1}) remains Hermitian only when
$L^\alpha_\bk$ and $N_\bk$ both vanish for all $\bk$. This requires $R_1=R_3=0$ and 
$R_4=-R_2$. Under these conditions the resistance tensor (\ref{rho}) becomes purely off-diagonal and
antisymmetric. This form signifies a purely transverse,
non-dissipative response -- an ``ideal Hall resistor''. 

It is important to note that the resistance tensor, Eq.\ (\ref{rho}), is not
invertible in this limit. 
As a consequence, the current response to applied voltages is ill-defined.  However, we can still achieve sensible results
by keeping a small non-zero dissipative components $R_1=R_3=R$. This causes the network
Hamiltonian to become non-Hermitian and
results in weak damping of the AC signal. Topological properties of the system should not be affected as we explicitly illustrate
below. Large non-Hermitian components could lead to new interesting
topological phases and will be discussed elsewhere.

We now focus on the approximately Hermitian limit and define the Hall parameter $R_H=R_4=-R_2$.
The bulk spectrum corresponding to the Hamiltonian (\ref{h1}) is illustrated in Fig.\
\ref{fig1}(c). It develops a gap 
$\Delta=4R_H\sqrt{C/L}$ measured in units of $\omega_0=1/\sqrt{LC}$ at
$\bk=0$. Since the term $M_{\mathbf{k}}$, responsible for the
gap formation, is odd under time reversal, we expect the gapped phase
to be topologically non-trivial. An explicit calculation indeed
indicates a non-zero Chern number $c=\sgn{R_H}$ for the negative
frequency band. Numerical calculation of the spectrum in a strip
geometry confirms the existence of a single chiral edge mode traversing
the gap, as shown in Fig.\ \ref{fig1}(d).

Experimental characterization of a finite size network can be given
through the two-point impedance measurement which is conveniently described by the circuit Green's function formalism. Here we give a brief review of the formalism while a more detailed discussion can be found e.g.\ in Ref.\ \onlinecite{Lee.2018}. 
To begin one writes the frequency-domain Kirchhoff law for current 
conservation in the matrix form 
\begin{equation}
0=\sum_{\br'} Y_{\br\br'}(\omega)V_{\br'} \,,
\end{equation}
which defines the admittance tensor $Y_{\br\br'}(\omega)$.
The eigenmode spectrum can be calculated from the condition 
$\text{det} {\bm Y}(\omega) = 0$, which yields results equivalent to Eq. (\ref{kir4}).
The circuit Green's function 
$G_{\br,\br'}(\omega)=\left[Y(\omega)^{-1}\right]_{\br\br'}$ 
describes the voltage response
of the network at point $\br$ to a driving current profile $I^{\rm drive}_{\br'}$
according to 
\begin{equation}\label{green1}
V^{\rm resp}_\br(\omega)=\sum_{\mathbf{r'}} G_{\br,\br'}(\omega)I^{\rm drive}_{\br'}(\omega) \,.
\end{equation}
In analogy to condensed matter systems, where the complete characterization
of a non-interacting system is contained in the two-point correlation function
$\langle \cT \psi(\br,t) \psi^\dagger(\br',t') \rangle$, full experimental knowledge of
$G_{\br,\br'}(\omega)$ provides a complete characterization of the
electrical circuit. We can therefore expect topologically non-trivial behavior to be evident
in a circuit's two-point impedance. 

In Fig.\ \ref{fig1}(e) we demonstrate this explicitly
by plotting the voltage profile $V^{\rm resp}_\br$ induced
by a current with frequency $\omega$ injected at the boundary of a $10\times 10$
network. As an example of possible dissipative dynamics we include a
non-zero $R$ component of the resistance tensor $\hat{R}$ and
quantify the strength of dissipation by a dimensionless parameter
$\epsilon=R/R_H$.  For the frequency inside the bulk bandgap the
signal is seen to propagate along the
boundary of the system and in one direction only, consistent with the
chiral nature of the gapless edge mode. Parameter $\epsilon$ clearly
controls the lengthscale over which the signal is damped.

\begin{figure}[tb]
  \centering
  \includegraphics[width=\columnwidth]{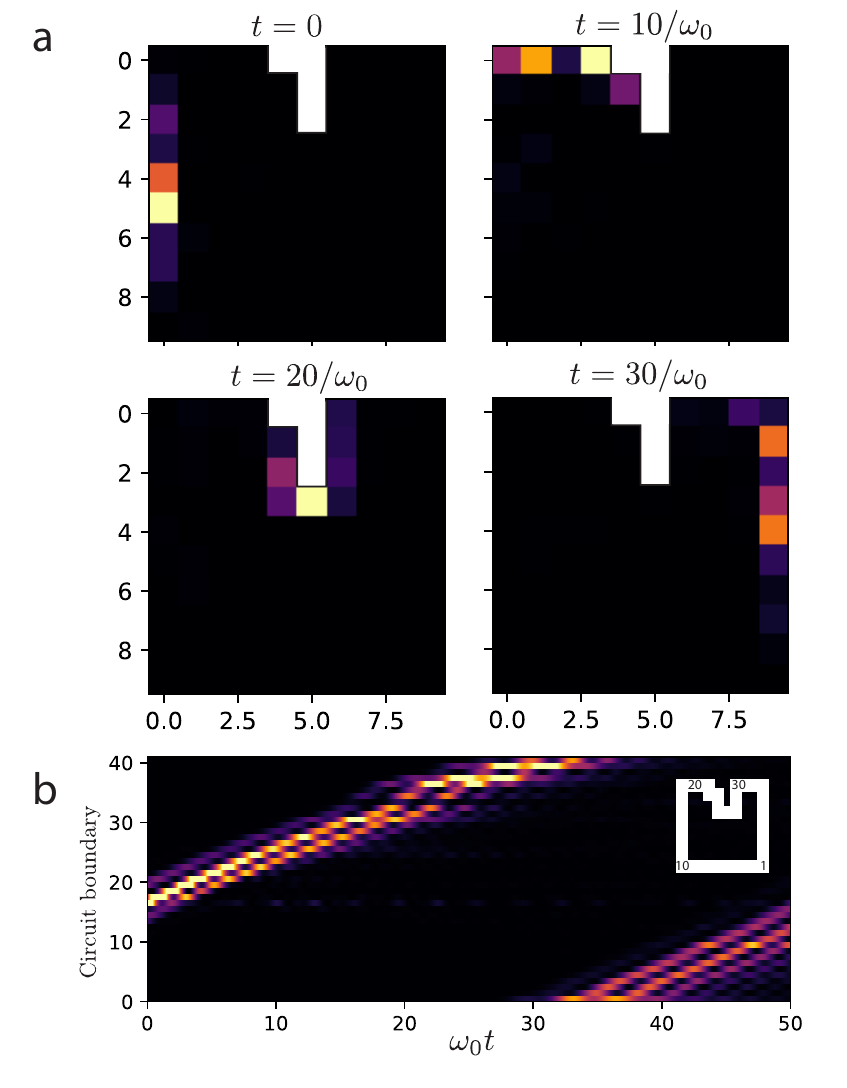}
  \caption{(a) Time evolution of a localized Gaussian wave packet of  frequency width $(\Delta\omega)/\omega_0=0.35$ excited at the boundary. The simulation models disorder by assuming a capacitor and inductor device tolerance of $30\%$. Colorscale corresponds to the weight of the wavefunction 
  on the circuit node. The signal travels along the boundary 
  and circumvents the boundary defect indicated in white. 
  (b) A plot of the current along the boundary sites as a function of time reveals the constant group velocity of the wave packet. }
  \label{fig:evo}
\end{figure}

Finally, we investigate the propagation of such signals in the time domain.
To this end we excite a Gaussian wave packet with the frequency width 
$(\Delta \omega)/\omega_0=0.35$ spatially localized around an edge site and
unitarily evolve it in time with the propagator $U=\exp (-i H t)$. 
The corresponding simulation for a non-dissipative network
with $\gamma=1$ and assuming $\pm30\%$  randomness in $L$ and $C$ values 
is shown in Fig. \ref{fig:evo}(a). 
The edge signal propagates unidirectionally along the circuit boundary, even in the presence of boundary defects.
A plot of the current at the network boundary as a function of time in Fig. \ref{fig:evo}(b) reveals approximately constant group velocity of the wave packet.

The circuit described above illustrates the mathematical 
correspondence between periodic RLC networks
and tight-binding Hamiltonians with non-trivial topology. Our approach
allows for the mapping of the
differential equations governing the RLC network onto a 
simple Bloch equation known in the condensed-matter literature
\cite{Weeks.2010}.  The non-trivial ingredient required to break
time reversal symmetry 
is the 5-terminal Hall element described by the resistance tensor, Eq. (\ref{rho}). 
However, as we will discuss in Sec. \ref{sec:implementation},  its
experimental realization is not straightforward. For this reason, we
may regard the above network as an instructive but unphysical toy
model. Below  we will describe two different network
architectures which have well-defined experimental implementations and
are only slightly more complex.

\subsection{Chern insulator on the square lattice}
\label{sec:model2}
\begin{figure*}
  \centering
  \includegraphics[width=17cm]{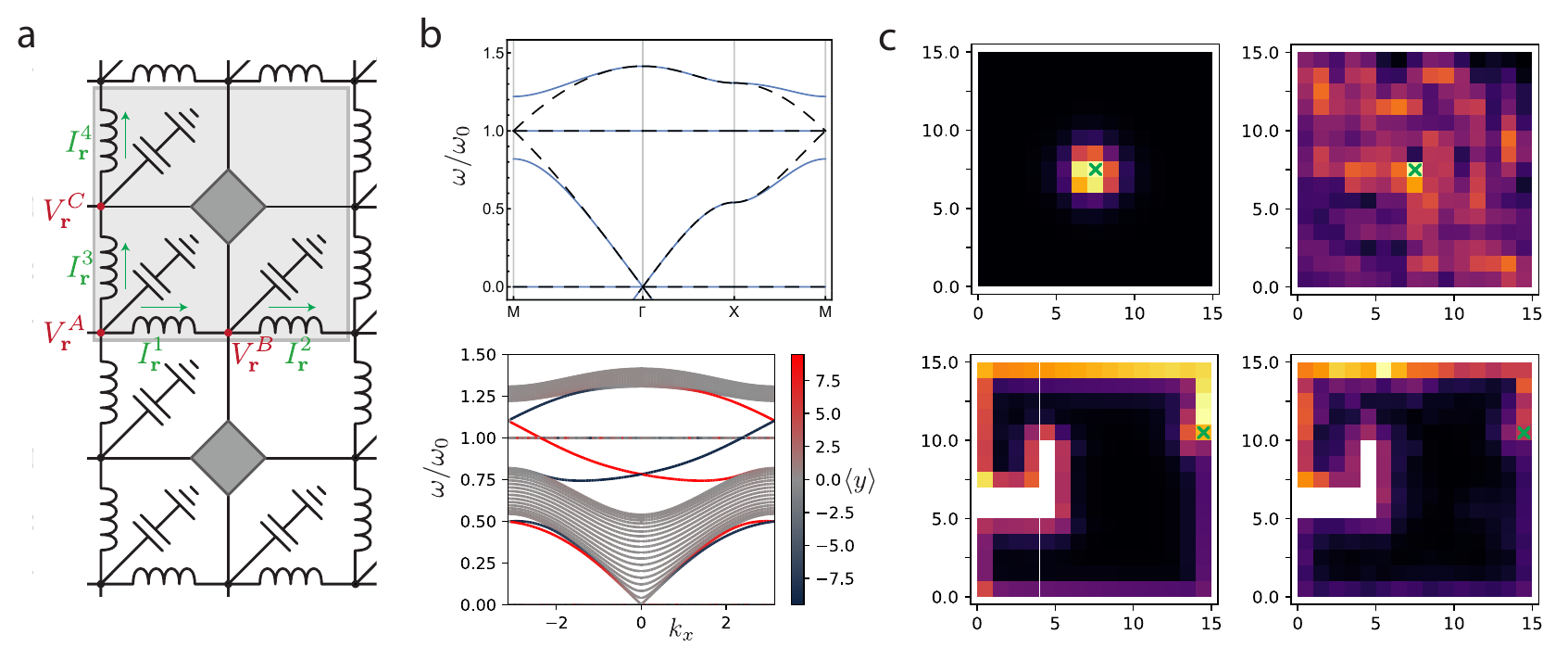}
  \caption{(a) Square RLC lattice network with 4-terminal Hall elements. 
  Voltage nodes (red) and currents (green) are labeled for the unit cell (grey background) at
  position $\br$. 
  (b) Eigenmode spectrum of the network for $g=1/\sqrt{2}$, with (solid lines) and without the Hall element (dashed lines).  The gap parameter is  $\gamma=\sqrt{C/L}R_H=5\sqrt{2}$ and we have defined an overall frequency scale $\omega_0=\sqrt{2/LC}$. 
  (c) Strip-diagonalization of the network with $g=1/\sqrt{2}$ and  $\gamma=5\sqrt{2}$.
  Colorscale
  indicates the average distance $\langle y\rangle$ measured from the center of
  the strip of the eigenstate belonging to the eigenvalue
  $\omega_k$.
  (d)  
  Voltage response $V^{\rm resp}_\br(\omega)$ of a $15\times 15$ circuit with
  bandstructure as in c 
  to current injected at green marked
  sites. Frequencies of the injected currents are denoted in plot titles. For all plots we assume a small resistance of the inductors $\varepsilon=0.005$ responsible for damping of the signal.
  Bottom panels include topological defect where white sites have been removed. For the bottom right panel we additionally model $17\%$ randomness in $L$, $C$, $R_H$, and $\varepsilon$ values.
  }
  \label{fig3}
\end{figure*}
Consider the network depicted in Fig.\ \ref{fig3}(a). It has a square
lattice symmetry and contains 4 inductors, 3 capacitors and one Hall
resistor per unit cell.
The Hall resistor is now in a 4-terminal configuration. 
We characterize it by the Hall admittance tensor $\hat{Y}$ that relates input currents to terminal
voltages via $I=\hat{Y} V$.
In its idealized version it is 
\begin{equation}
	\left(
	\begin{array}{c}
		I_1 \\ I_2 \\I_3 \\I_4
	\end{array}
	\right)
	=
	\frac{1}{R_H}
	\left(
	\begin{array}{cccc}
		0 & -1 & 0 & 1 \\
		1 & 0 & -1 & 0\\
		0 & 1 & 0 & -1\\
		-1 & 0 & 1 & 0
	\end{array}
	\right)
	\left(
	\begin{array}{c}
		V_1 \\ V_2 \\V_3 \\V_4
	\end{array}
	\right)
	\,.
	\label{eqn:4term_y}
\end{equation}
Currents and voltages are labeled as shown previously in Fig.\ \ref{fig1}(b) with the difference that no central terminal exists. We note that $\hat{Y}$ has rank 2 and is therefore not invertible. We can reduce (\ref{eqn:4term_y}) to a set
of two linearly independent equations by realizing that it
conserves current for pairs of opposing terminals, that is, for any
voltage input the currents satisfy $I_1=-I_3$ and $I_2=-I_4$.
In electrical circuit theory this is known as the port condition. Two opposing terminals
define a port. A full description of the Hall element is then
achieved in terms of two currents
through the ports, $I_1$ and $I_2$, and two voltages across the ports,
$V_1-V_3$ and $V_2-V_4$. The corresponding
resistance tensor is
\begin{equation}
	\hat{R}=\hat{Y}^{-1}=
    \begin{pmatrix}
        0 & R_H  \\
        -R_H & 0
    \end{pmatrix}
    \,.
    \label{eqn:r2port}
\end{equation}
We note that the circuit element corresponding to the above resistance matrix 
is in fact well known in electrical engineering literature as the gyrator \cite{tellegen1948gyrator}. This device,
together with the resistor and the capacitor, defines a basis 
of linear circuit elements. All other network elements can be composed
from the  aforementioned three.

The degrees of freedom describing the network in Fig.\ \ref{fig3}(a)
can be chosen as three voltages on the capacitors and four currents flowing
through the inductors, forming a 7-component vector
$\Psi_\br=(V^A_\br,V^B_\br,V^C_\br, I^1_\br,I^2_\br,I^3_\br,I^4_\br)^T$. 
To preserve the 4-fold rotation symmetry of the network, we take
capacitances on B and C sublattices to be equal, $C_B=C_C=C$, and further set
$C_A=C/g^2$ with $g$ a dimensionless parameter. All inductors have 
inductance $L$.

The corresponding Bloch Hamiltonian follows from 
current conservation for all nodes and Kirchhoff's second law for the potential
difference between two nodes connected through an inductor. 
It can be represented as a $7\times 7$ matrix of the form
\begin{equation}
  \label{h2}
H_\bk =
{\frac{1}{\sqrt{LC}}}
\begin{pmatrix} 
M_\bk & P_\bk  \\
P_\bk^\dagger  & \hat{0}
\end{pmatrix},
\end{equation}
where $\hat{0}$ is a $4\times 4$ matrix with all elements
zero and $P_\bk$ denotes the $4\times 3$ matrix
\begin{equation}\label{h3}
P_\bk =
i
\begin{pmatrix} 
-g & g e^{-i k_x} & -g & ge^{-i k_y}  \\
1 & -1 & 0 & 0 \\
0 & 0 & 1 & -1
\end{pmatrix}.
\end{equation}
The $3\times 3$ matrix $M_\bk$ contains time reversal breaking terms due to presence of the
the Hall element,
\begin{equation}\label{h4}
M_\bk =
\begin{pmatrix} 
0 & 0 & 0  \\
0 & 0 & m_\bk \\
0 & m_\bk^* & 0 
\end{pmatrix},
\end{equation}
with $m_\bk={\frac{i }{R_H}}\sqrt{\frac{L}{C}}(1-e^{i k_x})(1-e^{-ik_y})$.

The mode spectrum of the circuit consists of 7 bands. 
Charge-conjugation symmetry $\cC$
constraints the bands to come in pairs of opposite frequency and the unpaired
band to be confined to $\omega_\bk=0$. In the absence of the Hall
resistor time reversal symmetry enforces degeneracies at $\bk=(0,0)$
and $(\pi,\pi)$ as follows
\begin{equation}\label{deg1}
\begin{split}
\omega_{(0,0)}&=\omega_0(0,\pm 0,\pm 1,\pm\sqrt{1+2g^2}), \\
\omega_{(\pi,\pi)}&=\omega_0(0,\pm 1,\pm 1,\pm\sqrt{2}g).
\end{split}
\end{equation}
Here, we have defined $\omega_0=\sqrt{2/LC}$. The Hall resistor breaks $\cT$ and splits the degeneracy at
$(\pi,\pi)$. The quadratic band crossing thus acquires a gap and the
two bands become topologically non-trivial with the Chern number $c=\pm{\rm sgn}(R_H)$. 
Since $M_{(0,0)}=0$ the degeneracy at the $\Gamma$ point
remains intact.  

For an arbitrary $g$ and $R_H$ one thus expects the network to realize
a Chern insulator. A situation of special interest occurs for $g=1/\sqrt{2}$. In the
absence of the Hall resistor three bands then touch at $(\pi,\pi)$ and the
middle band is completely flat, Fig. \ref{fig3}(b). The Hall resistor
separates the three bands and makes the top and bottom bands
topological with Chern number $c=\pm{\rm sgn}(R_H)$. The flat band remains
trivial with $c=0$.
This is confirmed by numerical diagonalization of Hamiltonian Eq.\ (\ref{h2}) on a strip geometry with
translational invariance along $\hat{x}$, shown in Fig. \ref{fig3}(b). We
clearly observe chiral edge modes.

We further analyze the admittance properties of the network by calculating the
circuit Green's function in a finite system and plotting the voltage
response to a current injected at a single node. For these
calculations, we assume that the inductors are weakly resistive and characterize their resistance $R_L$ by a parameter $\varepsilon=R_L/R_H$. The resulting Hamiltonian becomes weakly non-Hermitian and the propagating waves are damped.

The upper left panel of Fig. \ref{fig3}(d) shows the response to a current of
frequency $\omega$ within the gap that is injected at a bulk site. The voltage
profile is localized around the node of injection. 
If we tune the frequency out of the bulk gap, the voltage signal propagates
through the whole circuit, independent of the point of injection, as shown in
the upper right panel. To demonstrate the topological
nature of the edge transport, we include a defect on the 
circuit's left
boundary and excite the edge mode of the circuit at an in-gap frequency (bottom panels). 
As expected the signal propagates around the defect by following the
distorted edge. This does not change qualitatively when we introduce
bulk disorder, which we model  
by including a $17\%$ randomness in $L$, $C$, $\varepsilon$, and $R_H$ values, larger than
typical tolerances of commercially available electronic components.

\subsection{Chern insulator on the honeycomb lattice}
\label{sec:model3}
\begin{figure}
  \centering
  \includegraphics{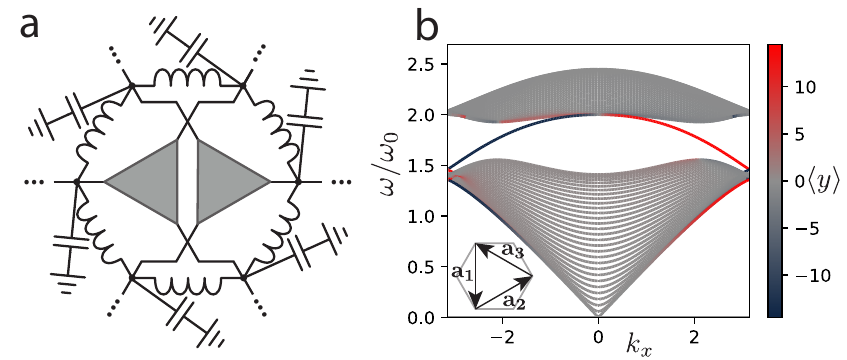}
  \caption{
  (a) Unit cell of a network realizing the honeycomb lattice model. 
  Three next-nearest neighbors within each plaquette connect 
  to a three-terminal Hall element. 
  (b)  Band structure of a strip with zig-zag termination 
  in $\hat{y}$-direction for $\gamma=R_H\sqrt{C/L}=10\sqrt{2}$ where $\omega_0=\sqrt{1/LC}$. Colorscale shows mean value of the distance of the corresponding eigenfunction
  from the center of the strip.
  }
  \label{fig4}
\end{figure}

The graph structure of RLC networks in principle allows for engineering arbitrary lattice models. Here, we
briefly discuss a circuit whose tight-binding analog is similar the Haldane model on the honeycomb
lattice \cite{Haldane.1988} which was historically the first
model realizing the Chern insulator in electronic systems.
A unit cell is schematically shown in Fig.\ \ref{fig4}(a). Each of the two sublattices of 
the honeycomb lattice contains a node that is connected to ground
through a capacitor $C$. 
Nearest-neighbor nodes are connected 
by inductors $L$. Second neighbors within a hexagonal plaquette each connect 
to a three-terminal Hall resistor.

The three-terminal Hall resistor is described by a three-fold rotation symmetric
resistance tensor whose idealized, non-dissipative form is defined by the relation
\begin{eqnarray}
  \begin{pmatrix}
  V_1 \\ V_2
  \end{pmatrix}
  =
  \left(
  \begin{array}{cc}
    0 & R_H \\
   -R_H & 0
  \end{array}
  \right) 
  \begin{pmatrix}
  I_1 \\ I_2
  \end{pmatrix}
  \,.
  \label{eqn:rh_haldane}
\end{eqnarray}
While above resistance tensor is formally equivalent to the matrix in Eq.\ 
(\ref{eqn:r2port}), the port condition does not apply for a triangular Hall element. 
Instead $\Hat{R}$ relates input currents at two of the three 
terminals to the corresponding terminal voltages. The potential at the third
terminal is set to zero and the corresponding  current is 
determined by current conservation.

The network in Fig.\ \ref{fig4}(a) has a tight-binding representation and topological structure
closely related to Haldane's celebrated lattice model of the Chern
insulator\cite{Haldane.1988}. The Bloch Hamiltonian is a $5\times 5$
matrix which takes the same form as Eq. \ref{h2}, where now
\begin{eqnarray*}
  M_{\mathbf{k}} = 
  \left(
  \begin{array}{cc}
     m_{\mathbf{k}} & 0 \\
    0 & - m_{\mathbf{k}}
  \end{array}
  \right),
\ \ 
  P_{\mathbf{k}} = i
  \left(
  \begin{array}{ccc}
    -1 & -1 & -1 \\
    1 & e^{i \mathbf{k \cdot a_1}} & e^{-i \mathbf{k \cdot a_2}}
  \end{array}
  \right)
  \,.
\end{eqnarray*}
Here, $m_{\mathbf{k}}=2 R_H^{-1} \sqrt{\frac{L}{C}}\sum_{\alpha=1}^3 \sin(\bk\cdot
\ba_\alpha) $ and $\ba_{i}$ are Bravais lattice vectors as shown in the inset of
Fig. \ref{fig3}(b).
Similar to graphene the band structure has a pair of Dirac points
located at the corners of the hexagonal Brillouin zone but they now
occur at non-zero frequency. The band crossings are protected by a
combination of $\cT$ and the lattice inversion symmetry. Inclusion of the Hall
resistors breaks $\cT$ and creates a Chern insulator.  
Numerical diagonalization of the Hamiltonian in the  strip geometry
confirms the existence of the chiral edge modes traversing the gap, Fig.\
\ref{fig4}(b).

\section{Hall resistor implementation}
\label{sec:implementation}
Non-trivial physics in the models studied in Sec.\ II above relies on Hall resistors
characterized by a transverse voltage response
to a longitudinal current that is odd under time reversal. 
Naively, the most straightforward realization of such devices exploits 
the classical (or quantum) Hall effect. 
Consider a 2D metal or semiconductor film in a perpendicular magnetic field.
In the strong field limit 
current flows along equipotential lines yielding dissipationless transport.
However, this picture neglects the influence of the boundaries of the Hall resistor. As we explicitly demonstrate in Appendix \ref{apdx:hall}, 
the boundaries give rise to a dissipative contact resistance\cite{Wick.1954}. It is 
therefore impractical to realize a circuit described by a Hermitian
eigenvalue problem in this manner. A possible workaround relies on capacitively coupling the Hall elements to the circuit\cite{Viola.2014}. We describe this effort in Appendix \ref{apdx:hall} and conclude that while it might work in principle a practical implementation is not straightforward.

On the other hand,
4-terminal Hall resistors that satisfy the port condition are well
known among electrical engineers  
as ``gyrators" and various other implementations have been 
conceived \cite{Hogan1952, Shenoi1965, Fabre1992, Castellanos-Beltran2009, Koch2010}. 
A notable example is the realization using operational amplifiers.
 Such gyrating circuits are discussed in standard
textbooks \cite{Chua.2015}.

\begin{figure}
	\centering
	\includegraphics[width=\columnwidth]{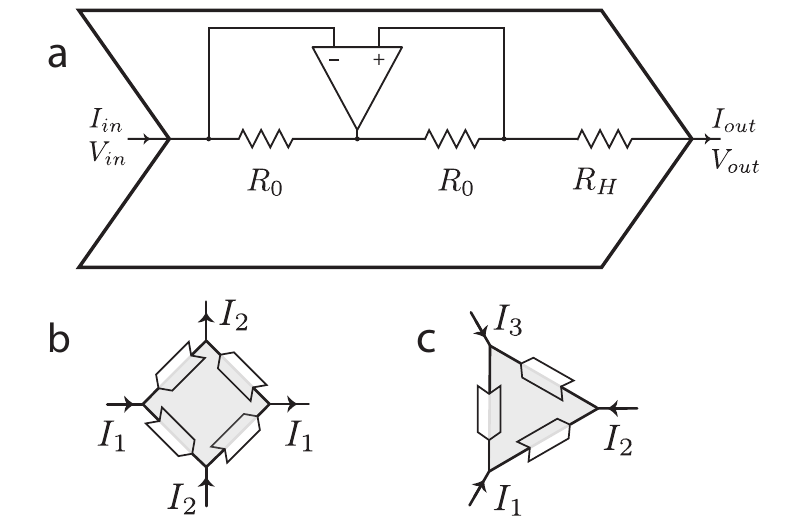}
	\caption{(a) Circuit element called "negative impedance converter", composed of three resistors and one operational 
		amplifier, introduced in Ref. \onlinecite{Hofmann.20180923}. It can be used to construct 
		an ideal Hall element in 2-port configuration (b) as implemented
		in the square lattice Chern insulating network discussed Sec. IIB, or  in a 3-terminal configuration (c) required
	in the honeycomb network of Sec.\ IIC.}
	\label{fig:opampcircuit}
\end{figure}

Here we describe a specific realization of the simulated ideal Hall
resistor inspired by the recent work of Hofmann \textit{et al}. \cite{Hofmann.20180923}.
It can be used in either 4- or 3-terminal configuration required for the
Chern insulating networks discussed in Secs.\ IIB and IIC, 
but not in the 5-terminal configuration.
Construction of the Hall element is based on the building block depicted in 
Fig.\ \ref{fig:opampcircuit}(a). It consists of an operational amplifier and three
resistors. A derivation of the corresponding admittance tensor is given
in Appendix A of Ref. \onlinecite{Hofmann.20180923}. It is
\begin{equation*}
	\left(
	\begin{array}{c}
		I_{\rm in} \\ I_{\rm out} 
	\end{array}
\right)
=
\frac{1}{R_H}
  \left(
  \begin{array}{cc}
    -1 & 1 \\
    1 & -1 
  \end{array}
  \right)
 \left(
 \begin{array}{c}
	 V_{\rm in} \\ V_{\rm out}
 \end{array}
 \right)
  \,,
\end{equation*}
where currents and voltages are defined as in Fig.\ 
\ref{fig:opampcircuit}(a). Remarkably, arranged in a two-port configuration
as depicted in Fig.\
\ref{fig:opampcircuit}(b) or three-terminal configuration in Fig.\ 
\ref{fig:opampcircuit}(c), these elements precisely realize the respective 
ideal Hall resistors required for our proposed
Chern insulator networks.

Operational amplifiers are commercially available at low cost and can
operate in a wide range of frequencies, voltages and power settings. Experimental realization of
the  Chern-insulating networks using the simulated Hall elements
depicted in Fig.\  \ref{fig:opampcircuit} should therefore be easily achievable.

\section{Summary and outlook}
In this work, we proposed periodic RLC networks that function as
Chern insulators for electromagnetic signals in a broad range of frequencies
tunable by adjusting the values of inductance $L$,
capacitance $C$ and Hall resistance $R_H$ of the circuit elements. 
The design is guided by exploiting an analogy between equations
governing the EM fields in periodic RLC
networks and tight-binding models for Majorana fermions which
are known to possess topologically non-trivial phases. 
Our approach maps the Kirchhoff's laws describing the network onto a Hermitian 
eigenvalue problem in the crystal momentum space where the eigenvalues
correspond to frequency modes of the network. Topological properties
of the network are then inferred transparently in direct analogy
to condensed matter Hamiltonians.

Explicitly, we have proposed three different network architectures
realizing Chern insulating phases for EM signals. The required
time reversal symmetry breaking is achieved by including Hall resistors which are non-reciprocal 
circuit elements also known in engineering literature as gyrators.
These may be implemented as capacitively contacted 
metallic or semiconductor films in an external magnetic field or
as simple circuits with resistors and off-the-shelf operational
amplifiers.  In the latter implementation the time
reversal symmetry is broken by the external source of power required
to operate the amplifiers. Nevertheless, due to the feedback structure
the operational amplifiers are operated in the linear response regime 
and the simulated Hall devices can be regarded as linear circuit elements.

Topological properties
of the networks proposed in this work are manifest in the chiral edge modes traversing 
the gap in the bulk spectrum. These edge modes give rise
to unidirectionally propagating voltage and current signals along the
network boundary. They are topologically protected by the bulk
topological invariant (the integer Chern number) and cannot be removed
by any deformation of the boundary. In addition, the edge modes are
robust against moderate amount of bulk disorder, as realized e.g.\ by a
random spread in the parameters characterizing the individual network elements.   

Chern insulating EM networks provide a highly tunable experimental environment. 
Scale invariance of Maxwell's equations allows for engineering of 
band gaps and  edge modes in a wide frequency range. Moreover, the
flexible graph nature of such networks removes
any restriction on dimensionality or locality. Consequently, 
exotic synthetic materials of 
arbitrary dimension and connectivity may be designed. In addition to possible
engineering applications, Chern insulating EM networks may be 
established as a teaching resource in university laboratory courses
and demonstrations.

\emph{Acknowledgments.}-- We thank Doug Bonn, Sarah
Burke, Gil Refael and Ronny Thomale for helpful discussions. Research described in this article was
supported by NSERC and by CIFAR. Software that facilitated this research was provided by CMC Microsystems. M.F.\ acknowledges the hospitality of The
Aspen Center for Physics and KITP Santa Barbara where part of this
work was completed. 

\emph{Note added.} -- When this work was essentially completed we
learned about a related proposal for a Chern-insulating LC network with
operational amplifiers serving as $\cT$-breaking elements \cite{Hofmann.20180923}. We have subsequently added a brief discussion in Sec. III.

\bibliography{RLC}

\appendix
\section{Relation to Majorana tight-binding models}
A general model for non-interacting Majorana fermions on a lattice is defined by a Hamiltonian of the form
\begin{equation}\label{maj1}
H=\sum_{ij} h_{ij}\gamma_i\gamma_j.    
\end{equation}
Here $\gamma_j$ are Majorana operators satisfying the canonical anticommutation relations 
\begin{equation}\label{maj2}
\{\gamma_i,\gamma_j\}=2\delta_{ij}, \ \ \ \gamma_j^\dagger =\gamma_j,    
\end{equation}
and $h_{ij}$ is an $N\times N$ matrix of tunneling amplitudes between lattice sites $i$ and $j$ (we assume the lattice has $N$ sites.). Hermiticity of $H$ together with relations \eqref{maj2} imply that $h_{ij}$ is purely imaginary and antisymmetric. We henceforth write it as $h_{ij}=it_{ij}$ where $t_{ij}$ is a {\em real} antisymmetric $N\times N$ matrix.

Solving the problem defined by Hamiltonian \eqref{maj1} is equivalent to diagonalizing matrix $\hat{h}$. Time evolution of an arbitrary state $\Phi$ is then governed by the corresponding Schr\"odinger equation 
\begin{equation}\label{maj3}
i\partial_t\Phi(t)=\hat{h}\Phi(t),
\end{equation}
where $\Phi(t)$ is regarded as an $N$-component state vector in the site basis. Using the property $\hat{h}=i\hat{t}$ we see that Eq.\ \eqref{maj3} becomes a purely real-valued wave equation which, therefore, admits purely real solutions $\Phi(t)$. 

It is this property of Majorana tight-binding models that motivates the connection to electrical networks which are also governed by real-valued wave equations, like \ Eq.\ \eqref{kir1}. As an example it is an easy exercise to show that Majorana fermions on the Lieb lattice with tunneling amplitudes illustrated in Fig.\ \ref{fig:boundaries}(a) are described by the same Hamiltonian as the RLC circuit discussed in Sec.\ IIA.

\section{Hall-resistor implementation by classical Hall effect}
\label{apdx:hall}
  
\subsection{Hall effect, galvanic coupling}
\begin{figure}
  \centering
  \includegraphics{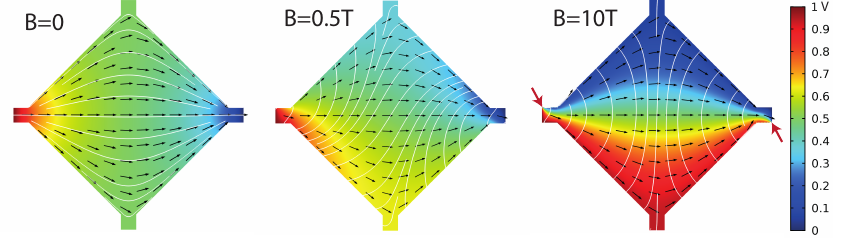}
  \caption{Microscopic simulation of the current and voltage distributions in
    a 2-dimensional Hall plate in a perpendicular magnetic field $B_{\bot}$
    with current density $\mathbf{j}$ driven by
    potential difference
    $V=\SI{1}{\volt}$ from left to right terminal. 
    White lines follow the electric field $\mathbf{E}$, black arrows 
    denote the direction of the 
    current flow $\mathbf{j}$. At zero field (left panel) $\mathbf{j}\parallel
    \mathbf{E}$ and there is no voltage drop $V_H$ between the top and the bottom terminal. For weak fields (middle)
    $\left|\mathbf{E}\right|\left|\mathbf{j}\right|>\mathbf{j}\cdot \mathbf{E} > 0$ and a small Hall voltage
    $V_H<V$ is observed. At high fields (right) $\mathbf{j} \bot
    \mathbf{E}$ and $V_H\simeq V$. In the infinite $B_\bot$-field limit the electric field
    diverges at the two contact points marked by red arrows.}
  \label{fig:comsol}
\end{figure}

We consider a metal or semiconductor film in a perpendicular magnetic field $B_\bot$.
In such a setting the microscopic current response 
is accurately described by Ohm's law, $\mathbf{j}=\sigma \mathbf{E}$, where the material's
conductivity takes the form \cite{ashcroft2010solid}
\begin{eqnarray}
  \sigma = \frac{\sigma_0}{1+\sigma_0^2 \cR_H^2 B^2}
  \left(
  \begin{array}{cc}
    1 & -\sigma_0 \cR_H B_{\bot} \\
    \sigma_0 \cR_H B_{\bot} & 1
  \end{array}
  \right) \,.
\end{eqnarray}
Here, $\sigma_0$ is the zero-field conductivity and
$\cR_H$ is the Hall coefficient. For $\sigma_0 \cR_H B \gg 1$, the
microscopic current response to a potential gradient
is predominantly transverse so one might think that a device
depicted in Fig.\ \ref{fig:comsol} could serve as a near-ideal Hall
resistor. As our simulations below illustrate, this unfortunately is
not the case because of the phenomenon of geometric magnetoresistance\cite{Wick.1954}. 

We use Comsol Multiphysics finite element software to numerically 
solve the current
conservation equation $\nabla \cdot \sigma \nabla V = 0$ for the 4-terminal geometry
depicted in Fig. \ref{fig:comsol}. We choose insulating boundary conditions for
the edges as well as
top and bottom terminals and drive a longitudinal current
$I$ by a voltage difference $V$ from the left to the right
terminal. The resulting potential distribution is plotted as a colorscale, 
electric field lines are white, and the current flow
is denoted by black arrows.
For $B_{\bot}=0$, the current flow is parallel to $\mathbf{E}$ and the potential
difference between top and bottom terminals is $V_H=0$. As one increases the
magnetic field, $\mathbf{j}$ and $\mathbf{E}$ span the Hall angle
$\theta_H$ and one measures a finite Hall voltage $V_H$. For constant current
flow, $V_H$ increases linearly with $B_{\bot}$. Naively, one could 
expect that $V_H \gg V$ for large enough field $B_{\bot}$. This is the necessary
condition for the realization of an ideal Hall element. 
However, as one can see in the right
panel of Fig.\ \ref{fig:comsol}, the Hall voltage saturates at $V_H=V$.

This effect is commonly known as two-terminal resistance
and may be interpreted as a geometrical magnetoresistance. For the
diamond
geometry it establishes a linear magnetic field dependence of the longitudinal
resistance. For constant current, $V_H$ and $V$ then show the same linear
behavior at high fields, precluding the desired limit $V_{H}\gg V$. In
fact, it has been shown that, on general grounds, $V_H\le V$ for arbitrarily shaped 3-, 4-, and 6-terminal geometry and arbitrary magnetic field\cite{Wick.1954}.

It may seem puzzling that a Hall element is dissipative
in the limit $\bj \bot \bE$. 
After all the dissipated power is $P=\int \bE \cdot d\bj$ and should
vanish when $\mathbf{j} \bot \mathbf{E}$.
But $P$ can still be non-zero
if the electric field strength diverges at some point in the sample. In fact, it is known that the two-terminal resistance arises at two points near the terminals where the boundary
conditions change from galvanic to electrically insulating. At these points, the electric field 
diverges. In our setup the points with divergent field strength are marked by red arrows in the rightmost panel of Fig.\ \ref{fig:comsol}.

\subsection{Hall effect, capacitive coupling}

\begin{figure}
	\centering
	\includegraphics[width=\columnwidth]{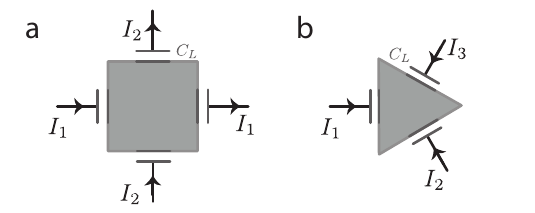}
	\caption{Illustration of capacitively contacted Hall elements in (a)
	four-terminal (two-port) and (b) three-terminal configuration. The
capacitance of each contact is $C_L$, directionality of currents is indicated by black arrows.}
	\label{fig:capel}
\end{figure}
Viola and DiVincenco \cite{Viola.2014} proposed an elegant way to circumvent
the problem of diverging electric fields outlined above. 
They showed that a near ideal Hall resistor can be achieved by
replacing galvanic contacts by  capacitive coupling to the
terminals. The resulting setup, illustrated in Fig.\ \ref{fig:capel}, 
yields solutions of the EM field equations that are well behaved on 
the whole resistor geometry. Explicitly, their resulting impedance tensor for 
a 2-port geometry Fig. \ref{fig:capel}(a) in the limit $\mathbf{j} \bot \mathbf{E}$ has
the form
\begin{equation}\label{imp}
    \hat{R}(\omega) = 
    R_H
    \begin{pmatrix}
    -i \cot{\left(\frac{1}{2}\omega C_L R_H\right)} & 1\\
    -1 & -i \cot{\left(\frac{1}{2}\omega C_L R_H\right)}
    \end{pmatrix}
    \,.
\end{equation}
Here, $C_L$ is the capacitance 
of a single contact, which are assumed to be the same for simplicity. The anti-symmetric structure of the tensor  implies that 
no energy is dissipated.
For a discrete set of perfect ``gyration'' frequencies
\begin{equation}\label{gyr}
\omega_n = \frac{\pi}{C_L R_H}(2n+1), \ \ \ n=0,1,\dots \,,
\end{equation}
 diagonal elements vanish and the
above tensor describes an ideal Hall resistor.

The impedance tensor of the capacitively contacted Hall element is 
intrinsically dependent on the drive frequency $\omega$ and this
dependence is fundamentally non-linear.
This prevents a description in terms of the  Bloch equation with a
simple frequency-independent Hamiltonian but one can still use the
circuit Green's function method to describe a periodic LC network with
these elements. 
We have checked numerically that the capacitively coupled Hall element
can indeed 
be used in network architectures  discussed in Secs.\ IIB and IIC and that it
produces Chern insulators, provided that the first gyration frequency
Eq.\ (\ref{gyr}) is tuned close to the frequency of the band crossing
where one wants to open a gap. This is not surprising given that at
$\omega=\omega_n$  the impedance tensor $\hat{R}(\omega)$ coincides
with the resistance tensor assumed in Secs.\ IIB and IIC, Eq.\
(\ref{eqn:r2port}). 

Simple physical description of the Viola-DiVincenco setup that gives
Eq.\ (\ref{imp})  relies on the dynamics of the magnetoplasmon edge mode in the Hall
effect device\cite{Viola.2014}. It will work equally well in the 4- and
3-terminal configuration, but the 5-terminal configuration that is
required for our toy model analyzed in Sec.\ IIA cannot be realized in this
manner.

\end{document}